\documentclass[aps,floatfix, twocolumn]{revtex4} 

\usepackage{graphicx}
\usepackage{dcolumn}
\usepackage{bm}
\usepackage{longtable}
 \usepackage{amsmath}
 \usepackage{amssymb}
 \usepackage{eurosym}
 \usepackage{subfigure}
\usepackage{color}
\usepackage{float}
\begin{document}

\title{Testing for voter rigging in small polling stations}

\author{Ra\'ul Jim\'enez$^{1}$, Manuel Hidalgo$^{2}$, Peter Klimek$^{3}$}
\email{peter.klimek@meduniwien.ac.at}
\affiliation{$^1$Department of Statistics, Universidad Carlos III de Madrid, 28903 Madrid, Spain\\ 
$^2$Department of Social Sciences, Universidad Carlos III de Madrid, 28903 Madrid, Spain\\
$^3$Section for Science of Complex Systems; CeMSIIS; Medical University of Vienna; 
Spitalgasse 23; A-1090; Austria
}

\begin{abstract} 
Since the 1970s there has been a large number of countries that combine formal democratic institutions with authoritarian practices.
Although in such countries the ruling elites may receive considerable voter support they often employ several manipulation tools to control election outcomes.
A common practice of these regimes is the coercion and mobilization of a significant amount of voters to guarantee the electoral victory.
This electoral irregularity is known as voter rigging, distinguishing it from vote rigging, which involves ballot stuffing or stealing.
Here we develop a statistical test to quantify to which extent the results of a particular election display traces of voter rigging.
Our key hypothesis is that small polling stations are more susceptible to voter rigging, because it is easier to identify opposing individuals, there are less eye witnesses, and supposedly less visits from election observers.
We devise a general statistical method for testing whether voting behavior in small polling stations is significantly different from the behavior of their neighbor stations in a way that is consistent with the widespread occurrence of voter rigging.
Based on a comparative analysis, the method enables to rule out whether observed differences in voting behavior might be explained by geographic heterogeneities in vote preferences.
We analyze 21 elections in ten different countries and find significant anomalies compatible with voter rigging in Russia from 2007-2011, in Venezuela from 2006-2013, and in Uganda 2011.
Particularly disturbing is the case of Venezuela where these distortions have been outcome-determinative in the 2013 presidential elections.
 \end{abstract}

\maketitle
\section{Introduction} 

Many elections around the world end in controversies  related to alleged frauds \cite{Norris2014};
even in mature democracies, such as the US and Canada, where voter suppression scandals have made the headlines\footnote{http://www.nbcnews.com/id/6242175/\#.VwPMgnqzEfA}$^,$\footnote{http://www.cbc.ca/news/politics/key-facts-in-canada-s-robocalls-controversy-1.2736659}. 
Yet while some  countries aim at ensuring trust in their electoral processes, by persecuting electoral malpractices, in others elections may take place under regimes that lack legitimacy.
Regrettably, in such broken societies, electoral irregularities may have very serious consequences ranging from social instabilities to deadly violence \cite{Norris2014b}.
There exist certain types of irregularities that seem to be characteristic of countries with extreme political polarization, with zones heavily controlled by only one political party.
For example, there are viral videos filmed during Zimbabwe elections that show electors that were forced to vote or allegedly bused under intimidation\footnote{We highlight the awarded video, made by a prison guard, that shows officers being forced to vote for President Robert Mugabe by superior officers (http://news.bbc.co.uk/2/hi/africa/7491077.stm). Other viral video shows Zimbabwean voters being bused from a rural district to a voting station in the capital towards the end of the day(https://www.enca.com/africa/viral-video-alludes-zim-vote-rigging)}.
It is an unsolved challenge to detect such abject electoral abnormalities by scrutinies of the vote counting or by audits of the electoral roll.
The reason being that behind every potential questionable vote there may exist a real elector whose vote officially counted.
Such abusive practices have been called {\it voter rigging}\footnote{http://www.news24.com/Africa/Zimbabwe/Zim-voter-rigging-captured-on-camera-20130801}, to signal unlawful and systematic harassments of the voters themselves (as opposed to distortions of the vote counts often referred to as ``vote rigging'').

Voter rigging is not exclusive of Zimbabwe.
Most of the complaints of the 2011 Russian elections came from state employees and students who said they were pressured by their supervisors/professors to vote for United Russia, which is a similar form of rigging\footnote{http://www.theguardian.com/world/2011/dec/04/russians-vote-national-parliamentary-elections}.
The 2013 Venezuelan presidential elections also seem to have been riddled with such voter rigging.
One of the main electoral observer groups  of these elections (ROE-AE) reported acts of violence or authority abuse during the election-day in 9.3\% of the observed voting centers.
ROE-AE denounced that in the 4.7\% of the cases there were infringements to the regulation on assisted voting\footnote{http://redobservacionelectoral.info/elecciones-presidenciales-2013-documentos-y-declaracion-entregados-al-cne/}.  These violations were used for compelling the voters to vote for the ruling party, according to the main opposition candidate\footnote{http://www.el-nacional.com/politica/Abusos-voto-asistido-empanan-resultados\_0\_175782533.html}.
The lack of integrity of the electoral process was such that the operations' headquarters of ROE-AE suffered a violent assault\footnote{http://www.dailymotion.com/video/xzn7kc\_observadores-electorales-atacados-en-las-elecciones-del-pasado-14-de-abril\_news}.
To our knowledge, this is the first time such an incident occurs in Venezuela. 
Another important electoral observation organization reported pressures on voters at different times during the day. 
In 15.1\% of the 391 observed voting centers, voter mobilization took place with public resources mainly belonging to government authorities and allied governorships and mayorships\footnote{http://www.oevenezolano.org/wp-content/uploads/2013/04/OEV-PRESIDENCIALES-2013-INFORME-FINAL-OEV.pdf}. 
In the opposition's appeal to the Venezuelan Supreme Court\footnote{http://esdata.info/pdf/201304-RecursoContenciosoElectoral.pdf},
where the outcome of the elections were contested, the inclusion of many small voting centers sensitive to voter rigging was mentioned.
Indeed, a point can be made that small centers might be particularly susceptible to voter rigging since it is easier to identify opposing individuals; 
these centers almost often lie in alleged pro-government areas, there are less eye witnesses; and they are visited less frequently be election observers.
However, beyond some undoubted proofs of voter rigging in a few voting centers, supported by amateur videos and observers' testimonies, 
there is no sound analysis about whether such electoral irregularities were isolated occurrences or if they happened on a larger scale, to the point of determining the winner of the election.

In this paper we provide a general method for testing irregularities of election outcomes due to voter rigging.
We hypothesize that the presence of voter rigging leads to a characteristic distortion in the election results that is especially discernible in small polling stations.
The method is applied to the results of 21 elections in ten different countries.
In particular we developed a statistical significance test that allows us to investigate whether the voting behavior in small polling stations is substantially different from large stations in a way that is consistent with the widespread occurrence of voter rigging.
The possibility of voter rigging in small stations is firmly rejected by our method in several elections in eight countries, including Venezuela before the current ruling party controlled the electoral power \cite{JH}. 
We observe a turning point in the size of election irregularities between 2004 and 2006 in Venezuela culminating in an outcome-determinative effect of voter rigging in the 2013 presidential elections.
In these elections, the voting behavior in small stations shows irregularities that can exceed the expected fluctuations of the results in the absence of voter rigging by a factor of almost ten.
The only other country where we observe anomalies of a comparable dimension is Russia, where the political landscape is dominated by United Russia.

The statistical detection of electoral fraud is not a new issue.
A number of scholars have been engaged in the study of this discipline during the past decade \cite{Levin2013}.
Among the most commonly-used statistical methods are those  based on Benford's laws \cite{Mebane2008, Pericchi2011} and other tests which also focus on the distribution of digits in vote counts \cite{Cantu2011, Sacco2012}.
In a separate category, 
we can group various tools developed for the detection of anomalies in the distribution of votes and voter turnout \cite{Myakgov2009, Klimek2012, Levin2009, rj2011}.
In a third category, we may include analyses based on exit poll \cite{Prado2011} and  other kinds of sampling data \cite{Hausmann2011, Enikolopova2012}.
Of interest are also studies that adopt a statistical mechanics approach to understand the statistical regularities in the vote counts \cite{Chatterjee2013}.
Particularly relevant are studies on the spatial correlation of turnout rates at the scale of municipalities \cite{Borghesi}.
Now, in order to incorporate spatial statistics for election fraud detection, we must consider more desegregated data sets, as it is well known that high levels of aggregation may mask electoral fraud \cite{Mebane2011}.
Furthermore, we have to analyze not only turnout rates but the complete election data set including vote counts.

\section{Data and Statistical Tools}

\subsection{Data}
\
For each election, we consider its vote counting at the finest available level of data collection that we denominate \emph{electoral unit}.
Such units may be known as electoral tables, wards, or precincts, according to the election under study.
We only use datasets that contain more than $1,000$ electoral units for which turnout, votes for the winner, and the number of electors in the unit, denoted by $n$, are all known and compatible. 
For each country, we also consider its partition into administrative divisions.
If different subdivisions are available, we use the smallest available territorial subdivision, seeking similarity among the units belonging to the same division.  
We refer to these divisions as \emph{electoral neighborhoods}.
They may correspond to different types of administrative entities depending on the country, such as departments, parishes, counties, districts or municipalities.
To ensure that the partition is sufficiently fine, we only consider datasets with more than $100$ electoral neighborhoods.
However, we require that each neighborhood has at least 10 electoral units in order to perform statistical tests of similarity among them.
Ten countries that we are aware of provide data that fulfill the above criteria.
We study 21 key elections of these countries listed in Table \ref{data}, together with the numbers of electoral units that meet our inclusion criteria, denoted by $N$, the average number of electors per unit of these electoral units, denoted by $\mu_n$, and the corresponding standard deviation, $\sigma_n$.
Across all the selected elections, $\mu_n$ only varies in one order of magnitude, which shows that the datasets have a comparable level of resolution.
In addition, $\sigma_n/\mu_n < 1$ for all elections, except  for Austria and France, which also shows also a comparable relative standard deviation of the number of electors per unit on 19 of the 21 case studies.

\begin{table}[tbp]
\caption{List of the 21 elections under study, with the numbers of electoral units that fulfill the inclusion criteria for our analysis, $N$. The average number of electors of these electoral units, $\mu_n$, is also given together with its standard deviation, $\sigma_n$.}
\label{data}
\begin{tabular}{l l l l l}
country & year & $N$ & $\mu_n$ & $\sigma_n$ \\
\hline
Austria & 2008 & 2,379 & 2,700 & 8,400\\ 
Canada & 2011 & 66,262 & 360 & 110\\  
Finland & 2011 & 2,352  & 1,900 & 1,600\\  
France & 2007 & 36,219 & 1,200 & 5,400\\  
Mexico & 2006 & 125,635 & 560 & 130\\  
Mexico & 2012 & 142,448  & 560 & 140\\  
Russia & 2003 & 95,181  & 1,100 & 900\\  
Russia & 2007 & 96,192 & 1,100 & 890\\  
Russia & 2011 & 95,057 & 1,100 & 860\\  
Russia & 2012 & 95,573 & 1,200 & 870\\  
South Africa & 2009 & 19,725 &  1,200 & 1,000\\  
Spain & 2008 & 59,346 &  570 & 190\\  
Spain & 2011 & 59,876 & 570 & 190\\  
Uganda & 2011 & 23,968 &  580 & 220\\  
Venezuela & 1998 & 20,026 &  550 & 170  \\  
Venezuela & 2000 & 10,340 & 1,100 & 700 \\  
Venezuela & 2004 & 23,562 &  590 & 140 \\  
Venezuela & 2006 & 32,336 &  480 & 110 \\  
Venezuela & 2012 & 38,853 &  480 & 110\\  
Venezuela & 2013 & 39,006 & 480 & 110 \\
Venezuela & 2015 & 40,546 & 480 & 110
\end{tabular}
\end{table}

\subsection{Standardized Election Fingerprints (SEF)} 
\
Let us denote by $t$ the turnout percentage per electoral unit and by   $vw$ the corresponding percentage of votes going to the winner.
The so called election fingerprint, namely a 2D-histogram for $t$ and $vw$, has proved to be a valuable tool for fraud detection \cite{Klimek2012} and is the starting point to introduce our methodology. 
The key idea for taking advantage of the election fingerprints is that they must fit approximately an uncorrelated bivariate normal distribution.
Nevertheless, this hypothesis may fail for many reasons. 
For instance, $vw$ might get inflated in a fraudulent way by adding to it votes from the non-voters (ballot stuffing).
In the election fingerprints this type of fraud will introduce a positive correlation between $t$ and $vw$.
Alternatively, $vw$ might be increased by adding votes from opposition parties (vote stealing), which would lead to an inflation of $vw$ in the election fingerprints, but not a simultaneous inflation of $t$.
There are also non-fraudulent mechanisms that can explain observed discrepancies between the election fingerprint of some countries and the uncorrelated bivariate Gaussian distribution.
For example, heterogeneity in the electoral population:
countries with two or more regions with different electoral behavior may correspond to Gaussian mixture models, which may provide multimodal fingerprints. This can happen if each region fits a normal distribution with different means and variances among the regions.
This appears to be the case of the 2011 Canadian elections where the fingerprint splits Quebec from English Canada \cite{Klimek2012}.
On the other hand, one may expect some sort of correlation between $t$ and $wv$,
especially if the voters of some electoral units are mobilized to support an option different from their first preferences in order to prevent an undesirable outcome (\emph{strategic vote}).
Strategic vote is common in several countries with proportional representation electoral systems \cite{Freden} and could explain the aspects of the Finnish and Austrian fingerprints, among others \cite{Klimek2012}.

\begin{figure}[tbp]
\begin{center}
 \includegraphics[width=80mm]{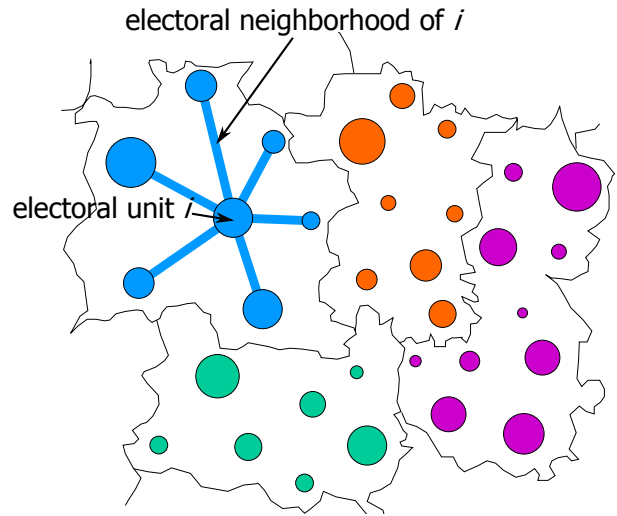}
\end{center}
 \caption{Schematic overview of how to compute the $Z$-scores. In this illustration we show electoral units (circles) with different electorate sizes $n$ (indicated by the size of the circles). The colors of the units correspond to different electoral neighborhoods. The neighborhood of $i$ is defined as all other units that lie in the same electoral neighborhood (blue lines). To correct for geographic heterogeneities in the data, the vote and turnout percentages of each unit $i$ are rescaled by their average value and standard deviation in the electoral neighborhood of $i$}
 \label{map}
\end{figure}

In order to provide an alternative forensic tool to the election fingerprints that is sturdier against the effects of the non-fraudulent scenarios discussed above,
we consider a stratified normalization of the percentages $t$ and $vw$ that we name the election $Z$-scores. 
Therefore we compare vote and turnout at a particular unit to the results of units in its neighborhood, see Figure \ref{map}.
Namely, the $Z$-scores of the electoral unit $i$ are
\begin{equation}
Z_t(i) = \frac{t(i) - \mu_{t}(i)}{\sigma_{t}(i)}\ \ \mbox{and} \ \  Z_{vw}(i) = \frac{vw(i) - \mu_{vw}(i)}{\sigma_{vw}(i)},
\label{zscores}
\end{equation}
where $\mu_t(i)$ and $\sigma_{t}(i)$ denote the average and standard deviation of $t$ over the units lying in the neighborhood of unit $i$
and $\mu_{vw}(i)$ and $\sigma_{vw}(i)$ the corresponding average and standard deviation of $vw$.
We will refer to the 2d histograms of the $Z_t$ and $Z_{vw}$ values for a given election as its {\it Standardized Election Fingerprint} (SEF).
For an easier accessible visualization, we will represent these 2d histograms also by smoothed level curves for the joint density of data points, see Materials and Methods section; different density levels are represented by proportional color intensities.

\subsection{Statistical test for voter rigging}
\
We now introduce a statistical test for the detection of voter rigging in small electoral units by developing a comparative election forensic tool based on SEFs and $Z$-scores of different countries.
But first, let us introduce some additional notation.
Let us index all elections in our data by the index $k$.
Denote by $S(k,p)$ the set of electoral units with less electors than the $p$-th percentile of the number of electors per the electoral unit (\emph{small units}) in election $k$.
Let $L(k,p)$ be the complementary set of electoral units (\emph{large units}).
In the following we only consider cases where $S(k,p)$ and $L(k,p)$ contain, both, at least ten elements.
We are interested in the detection of outlier elections in the sense that the joint distribution of the $Z$-scores in $S(k,p)$ units (SEF of small units) differs significantly from the distribution of the $Z$-scores in $L(k,p)$ units (SEF of large units).
This matter leads to a bivariate two-sample problem, a statistical problem of permanent interest {\it per se}.
As it is well known, the available tests for these problems usually depend on the kind of problem under consideration \cite{Franz}. 
Our approach is based on a simple comparison between the SEFs of small and large units.
Without dismissing the potential usefulness of other discrepancy measures between distributions, we will address the comparison between the SEFs of small and large units by considering the distance between their centers.
For simplicity, we will consider a standardized Euclidean distance between the centers of these distributions.
Estimating the coordinates of the centers by the median of the related $Z$-scores, this is
\begin{equation}
m_t^S(k,p) = \mbox{median}[Z_t(i), i\in S(k,p)],
\label{median}
\end{equation}
and similarly $m_{vw}^S(k,p)$, $m_t^L(k,p)$ and $m_{vw}^L(k,p)$,
the plugin estimator of the distance between the centers is 
\begin{widetext}
\begin{equation}
D_k(p) = \sqrt{[m_t^S(k,p)-m_t^L(k,p)]^2 + [m_{vw}^S(k,p) - m_{vw}^L(k,p)]^2}.
\label{distance}
\end{equation}
\end{widetext}

A central idea of the statistical test for voter rigging is to compare values of the distance $D_k(p)$ for election $k$ to its expectation from a set of different, trustworthy elections.
To this end we compute the values of $D_k(p)$ over all considered elections and identify the outliers in this set using the modified Thompson Tau test at a given confidence level $\alpha$, see Materials and Methods.
This test is applied to a wide range of choices of $p$, $p \in \{0.5, 1, 1.5, 2, \dots, 90 \}$.
We obtain a reference set of (trustworthy) elections, $R$, by considering all elections that are {\it not} classified as outliers for at least $(1-\alpha)\cdot100$\% of size thresholds $p$.
As a measure for the effect size for a given election we rescale its value of $D_k(p)$ by the mean and standard deviation of the corresponding distances of elections contained in $R$.
Hence, we consider the standardized Euclidean distance
\begin{equation}
\delta_k(p) =  \frac{ D_k(p) -  \mbox{mean}[D_l(p), l\in R]}{\mbox{std}[D_l(p), l\in R]},
 \label{deltaeq}
 \end{equation}
with $ \mbox{mean}[D_l(p), l\in R]$ and $\mbox{std}[D_l(p), l\in R]$ being the mean and standard deviation taken over trustworthy elections.
Thus, $\delta_k(p)$ values far from zero imply atypical distances, unexpected in free and fair elections. 
In particular we can provide a {\it rejection region} at a given significance level for $\delta_k(p)$ for the hypothesis of voter rigging in election $k$ by considering the rejection region of the corresponding modified Thompson Tau test.
If $\delta_k(p)$ lies outside of this region and, additionally, the center of the SEF for election $k$ of small units is inside the upper right region of the plot, the outcome of election $k$ is compatible with the hypothesis of large-scale voter rigging in small electoral units.

\section{Results}

\begin{figure*}[tbp]
\begin{center}
 \includegraphics[width = 0.9\textwidth, keepaspectratio = true]{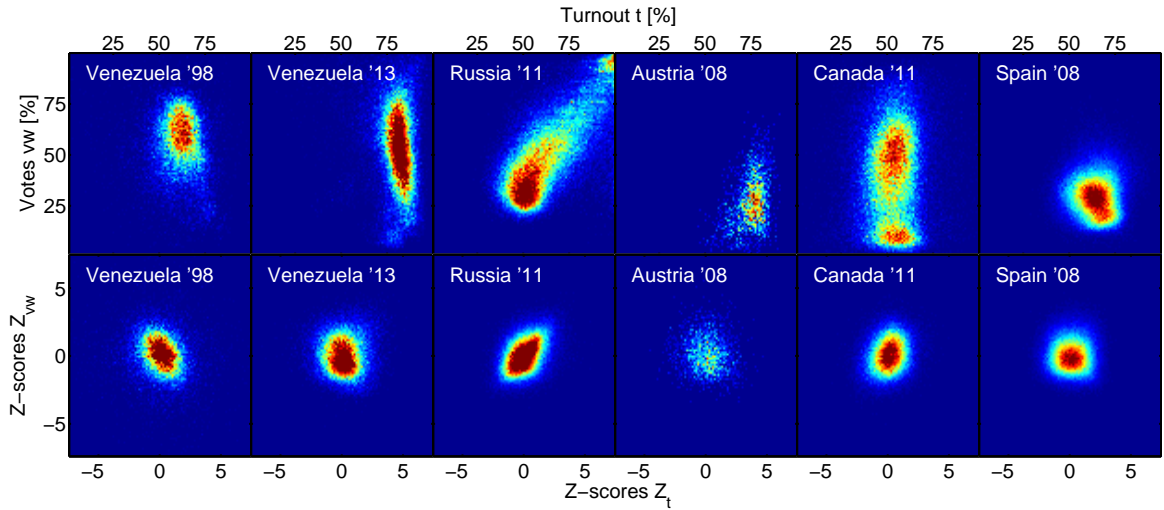}
\end{center}
 \caption{The SEFs (bottom row) are compared to the election fingerprints (top row) as introduced in \cite{Klimek2012} for a selected set of elections, namely (from left to right) Venezuela 1998 and 2013, Russia 2011, Austria 2008, Canada 2011, and Spain 2008. The bimodality in the fingerprint of Canada and the ``smearing out'' in Austria disappear in the SEF.}
 \label{SEFs}
\end{figure*}

\begin{figure}[tbp]
\begin{center}
 \includegraphics[width = 0.45\textwidth, keepaspectratio = true]{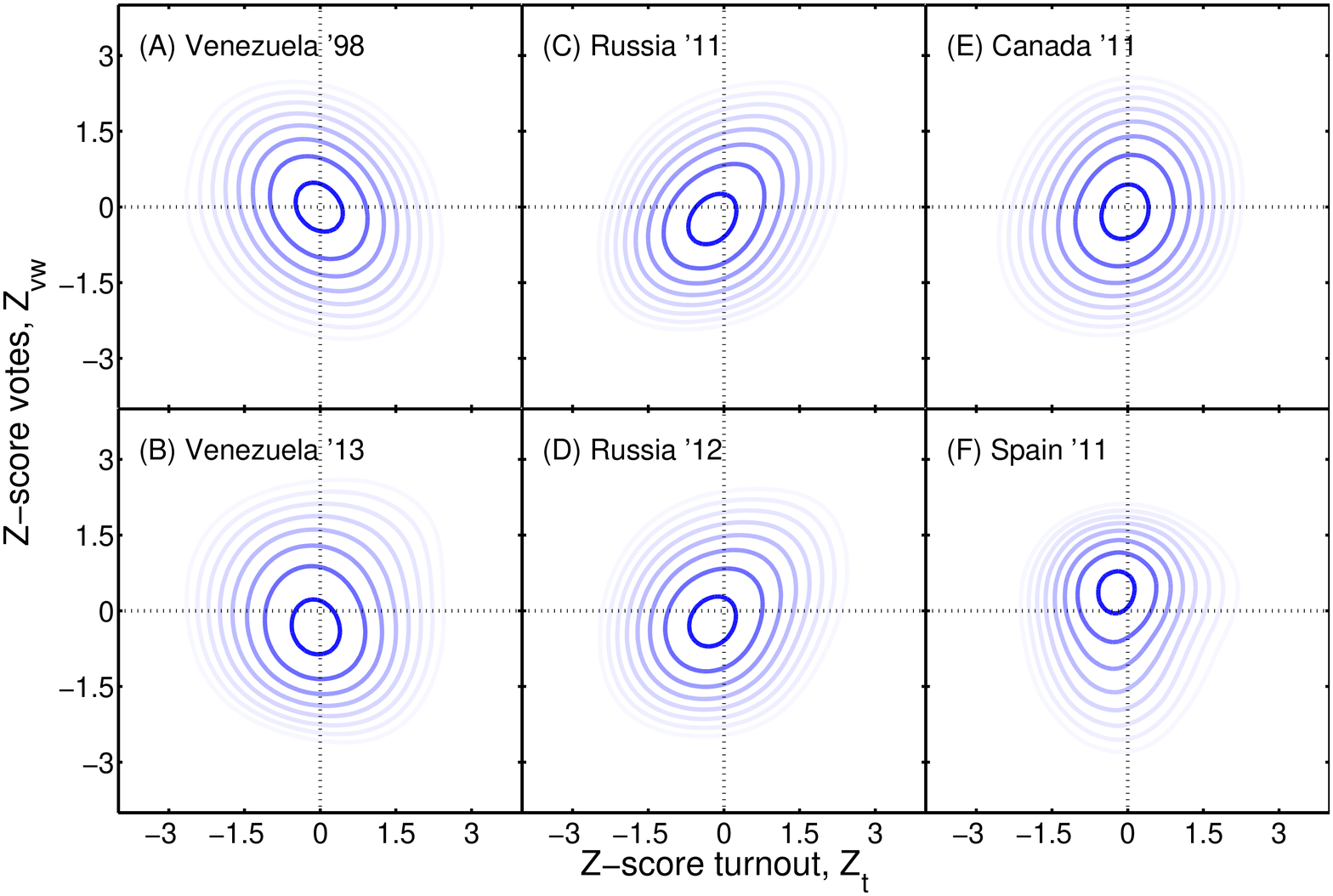}
\end{center}
 \caption{The SEFs of (A) Venezuela 1998 and (B) 2013, (C) Russia 2011, (D) Russia 2012, (E) Canada 2011, and (F) Spain 2011 are shown using contour representations for the densities of data points. The SEFs appear to be elliptically symmetric.}
 \label{symm}
\end{figure}

Figure \ref{SEFs} shows the election fingerprints as introduced in \cite{Klimek2012} and the SEFs of several elections (Venezuela 1998 and 2013, Russia 2011, Austria 2008, Canada 2011, and Spain 2008).
Note that the stratified standardization in the SEFs corrects multimodality (Canada) and heterogeneous voter mobilization (Austria) that has been observed in previous versions of election fingerprints.
Nonetheless, there is no reason to believe that SEFs should fit an uncorrelated bivariate normal model, such as it has been suggested for the fingerprints.
The single presence of strategic voting may introduce some correlation between $Z_t$ and $Z_{vw}$,
because more turnout may be associated to mobilizations for or against the winner.
If the electoral neighborhoods are properly chosen, making them homogeneous, 
asymptotic arguments can be invoked to argue that the empirical marginal distributions of the $Z$-scores should be approximately a standard normal distribution, although with more extreme values than the ones expected for a Gaussian sample \cite{rj2011, JH}.
Therefore, the only claim that we can assert is that the joint distribution should be unimodal, centered on the origin, and roughly supported on a high confidence normal area.
Additionally, we expect a particular symmetry for the SEF of an election where fraud is dismissed.
This becomes apparent in a contour visualization of the 2d histograms over all electoral units for several elections, see Figure \ref{symm}.
The SEFs of these elections appears to be elliptically symmetric.

\begin{figure}[tbp]
\begin{center}
 \includegraphics[width = 0.45\textwidth, keepaspectratio = true]{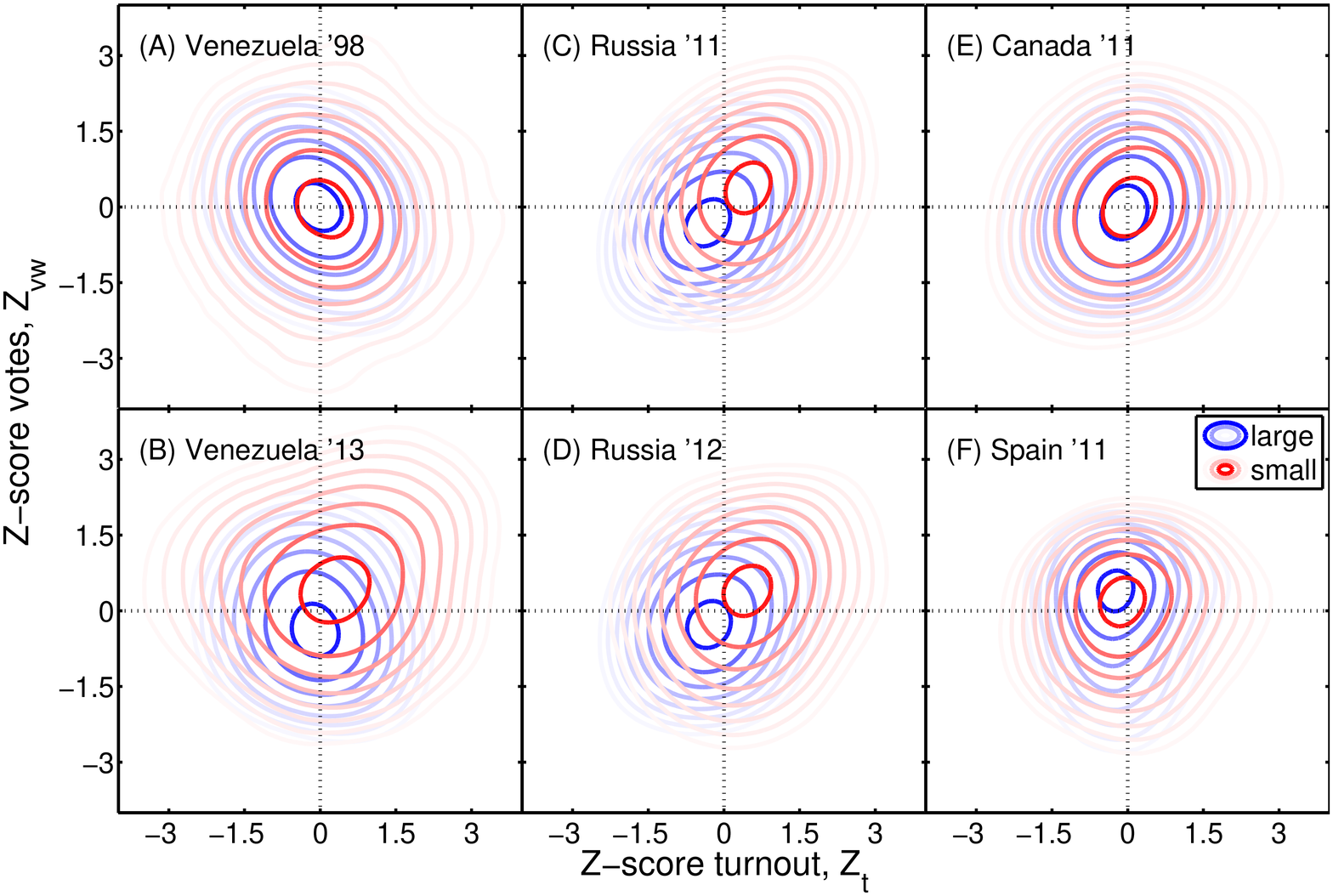}
\end{center}
 \caption{Comparison of the SEFs of large (blue) and small (red) electoral units for (A) Venezuela 1998 and (B) 2013, (C) Russia 2011, (D) Russia 2012, (E) Canada 2011, and (F) Spain 2011. The centers of the distributions of small and large centers co-incide for Venezuela 1998, Canada 2011, and Spain 2011. However, there is a clear discrepancy in the SEFs of small and large centers in Venezuela 2013 and Russia 2011 and 2012.  The rescaled turnout and the votes for the winner are substantially larger in small centers for these elections. This is clear evidence that the election outcomes in these small centers show systematic distortions.}
 \label{shift}
\end{figure}

\begin{figure}[tbp]
\begin{center}
 \includegraphics[width = 0.5\textwidth, keepaspectratio = true]{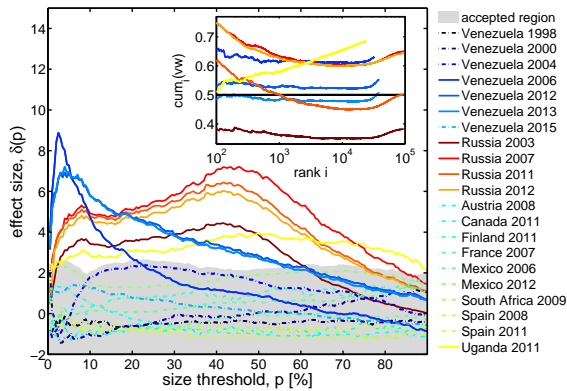}
\end{center}
 \caption{Results of the statistical test for voter rigging, quantified by the effect size $\delta(p)$, are shown for 21 different elections. The gray region contains the elections with no significant differences between small and large electoral units. These elections are also shown as dash-dotted lines. Solid lines show elections that are compatible with the voter-rigging-in-small-centers hypothesis. These include the elections in Russia, Uganda as well as Venezuela from 2006-2013, where the strongest effects are observed. In the inset we show a different visualizations of the atypical results observed from Russia, Uganda and Venezuela. In there, the electoral units are sorted in a descending way according to their number of electors and the percentage of votes for the winner is computed using only units up to the given rank on the x-axis (in logarithmic scale).
For Venezuela 2013 it is only the addition of small units that pushes the results to determine the winner.}
 \label{delta}
\end{figure}

Figure \ref{shift} shows visualizations of the SEFs for small and large electoral units, by using contour lines.
In this figure, small (large) units are those that have an electorate size below (above) the $p=20$ percentile.
Similar results can be observed for all other reasonable choices of $p$.
To rule out that our results are not driven by few small electoral units of questioned elections with atypical vote counting, we perform a simple outlier-removal procedure of $Z$-scores based on the observed elliptical symmetry (see Materials and Methods).
For Venezuela 1998, Canada 2011 and Spain 2011 the centers of the SEFs for small and large units coincide and the shapes of the distributions are hard to distinguish.
This is not the case for Venezuela 2013 and  Russia (2011 and 2012).
The $Z$-scores of these elections are substantially shifted towards the upper right regions of the plot for small units.
This means higher turnout values and larger numbers of votes for the winner in small electoral units as observed in their direct neighborhood,
a clear evidence of a systematic distortion of the election outcomes in these units that is consistent with the effects of voter rigging in such places.
In order to systematically quantify these distortions, we consider as effect size the distance $\delta(p)$, which gives the number of standard deviations by which the centers of small and large centers are displaced from each others as measured over a set of non-outlier elections.
Figure \ref{delta} shows results for $\delta(p)$ for the 21 elections given in Table \ref{data}.
The gray region in Figure \ref{delta}, the ``accepted region'', contains the results for all elections that are not classified as outliers according to the majority of outcomes of the modified Thompson Tau test.
Elections that lie outside of this region show results that are compatible with the assumption of widespread voter rigging in small electoral units.
We observe the strongest effects in Venezuela between 2006 and 2013, with $\delta(p)$ values that almost reach a factor of 10 for size thresholds around the 5th percentile.
Nevertheless, the values of $\delta(p)$ clearly lie outside of the accepted  region for a wide range of choices of $p$ 
Intriguingly, such strong deviations are totally absent from the Venezuelan data in prior elections as well as in the last Venezuelan parliamentary election, 2015.
The Russian elections between 2007 and 2012 also show significant deviations with $\delta(p)$ between 4 and 7.
Significant results are also found for Russian 2003 and Uganda, however with substantially smaller effect sizes with values between 3 and 4.
Certainly, these anomalous results for Russia, Uganda and Venezuela may be traces of voter rigging in small electoral units and raise serious doubts regarding the integrity of the related elections. 
Note that the accepted region in Figure \ref{delta} encapsulates elections that show deviations below three standard deviations, in consistency with a confidence level of $\alpha=0.95$.
The deviations in Russia from 2007--2012 and Venezuela 2006--2013 therefore indicate truly extreme events.

A different visualization of the atypical results reported for Russia, Uganda and Venezuela from 2006--2013 is provided in the inset in Figure \ref{delta}.
Here, the electoral units are sorted according to their number of electors in a descending way.
We then compute the percentages of votes for the winners over all units up to the given rank $i$ (i.e. above the corresponding number of electors).
This number is denoted $cum_i(vw)$.
If the voting behavior in the units is independent from the size of the unit, we expect a slope of zero for $cum_i(vw)$ for high ranks $i$.
This means that the addition of increasingly small units does not change the overall results of the elections.
For Uganda we observe a logarithmic increase of $cum_i(vw)$ from relatively large electoral units.
We observe a different pattern for Russia and Venezuela, where $cum_i(vw)$ clearly increases at the smallest units.
This means that the addition of these small units has a substantial impact on the election outcomes.
For Russia 2011 and Venezuela 2013  we note that it is indeed the contributions from the very small units that pushes the total number of votes to the barrier of 50\%.
In the Venezuelan case, where the current president (Nicol\'as Maduro) was elected by a plurality voting system, the systematic distortion of voting behavior in small electoral units was outcome-determinative. 

\section{Discussion}

In this article we develop a method for testing statistical anomalies of election results in small polling stations attributable to voter rigging. 
In particular, we devise a comparative tool based on Standardized Election Fingerprints (SEF) of different countries. 
Our analysis of twenty-one national elections in ten countries shows significant impacts in Venezuela 2006-2013, Russia 2007-2012 and, to a lesser extent, in Russia 2003 and Uganda 2011.
Particularly, traces of voter rigging are outcome-determinative in one case: the 2013 Venezuelan presidential elections.

In Russia and Venezuela from 1999 onwards with the assumption of power by Vladimir Putin and Hugo Ch\'avez, respectively, the dominant elites have progressively established an authoritarian competitive regime that combines formal democratic institutions with authoritarian practices. 
The ruling elites hold regular elections and tolerate some degree of freedom and competition for power, but on an uneven level playing field where  it is very difficult  for the opposition to win \cite{Levitsky}.
In fact, the ruling party has always won in Russia and in Venezuela, of the many elections held at the national level, the \emph{Chavismo} has only lost a constitutional referendum in 2007 and, more recently, the parliamentary elections of 2015. 
To stay in power, the elites have resorted to a mix of tools and practices to control the elections \cite{Schedler}.
Although the types and intensity of irregularities have varied over time, these include the banning of parties and disqualification of candidates, the abuse of state resources, hindrances and restrictions on  free press with a clear predisposition towards the officialism in  public media, biased electoral authority, no reliable electoral register, unfair electoral rules, and electoral fraud.  
These two hybrid regimes have turned more authoritarian in recent years, but, according to Freedom House, while Venezuela remains partly free, Russia is rated as not free since 2006\footnote{https://freedomhouse.org/report-types/freedom-world}.

The integrity of the Venezuelan electoral system has been questioned since the holding of the presidential recall referendum in 2004 \cite{JH}. 
An irregularity associated with our research has been the political and labor discrimination practices developed by the officialism against many citizens for signing a form to activate the recall referendum against then President Ch\'avez. 
Shortly afterwards, a list with the signatures was made public (Tasc\'on list) and was later refined with personal data, including benefits from any social mission (Maisanta list) \cite{MH2009}. 
In the public sphere, such practice created a climate of intimidation and mobilization of the vote for the ruling party. 
Despite some complaints in the media and reports by human rights organizations during the following years, this discrimination seems to have lost momentum after the period 2005-2006. 
Yet the mobilization of employees continued. 
In fact, the massive growth of highly clientelistic public employment enabled the officialism to mobilize employees at times with warnings and threats should they not support the Chavista government. 
However, the acute socio-economic crisis of recent years has made more difficult for officialism to mobilize voters but with its victories no mayor consequences were observed. 
That changed when the ruling party lost the parliamentary elections of 2015. 
Public employees have denounced to the media intimidation and verbal aggressions by their superiors\footnote{http://www.el-nacional.com/economia/Trabajadores-publicos-denuncian-agresiones-verbales\_0\_753524869.html}.
Additionally, in Venezuela other tactics to influence/pressure the voter in the last elections have become predominant. 
Firstly, through assisted voting, which may be associated with voter coercion. 
It was detected in 6,3\% of polling stations observed in 2012, 4,7\% in 2013, and up to 6\% in 2015, mainly on citizens that were pressured to vote for candidates of the ruling party\footnote{https://www.iidh.ed.cr/IIDH/media/3651/informe-final-capel-2015.pdf}. 
Secondly, the growth of small centers in the last decade. 
The electoral organism (CNE) has justified its policy  on the grounds of  the need to decentralize large centers and increase the number of centers in rural areas. 
However, these centers are more prone to irregularities and acts of intimidation/violence on election-days. 
In these small centers votes are mainly from citizens dependent on government social programs which make them very vulnerable to the modus operandi of the official machinery. 
In extremely competitive elections, such as the 2013 presidential elections, a manipulation in these centers may be critical to win the majority. 
What happened in the 2015 parliamentary elections? 
The types of irregularities reported two years before had less impact. 
In addition to lower mobilization of chavistas, largely due by the country's crisis, two factors seem to have played an important role: the deployment by opposition forces of activists and students, particularly in areas controlled by chavismo and therefore more vulnerable to possible fraud, 
and the institutional role played by the Armed Forces in an election more constrained by a stronger monitoring of the international  community.

In Russia, on the other hand, elections have not reached the minimum  standards to be considered democratic.
Pressure on voters is clearly apparent.
In the parliamentary  elections of 2003, there is evidence of pressure on several thousands of workers who were instructed by their employers to request absentee ballots and who were subsequently bused the day of the election so that their vote could be monitored  in previously designated centers under threat,  in some cases of job loss\footnote{http://www.osce.org/odihr/elections/russia/21482?download=true}.
In the parliamentary  2007 elections, there were allegations of threats against voters, of misuse of absentee ballots, and of voters being bused to designated centers\footnote{http://assembly.coe.int/nw/xml/XRef/X2H-Xref-ViewHTML.asp?FileID=11810\&lang=EN}.
In the parliamentary  elections of 2011, public officials were asked to sign letters of support for the ruling party. Owners of large companies also pressured their employees, instructed by local authorities to vote for United Russia\footnote{http://www.osce.org/odihr/elections/86959?download=true.}.
Like in the 2011 elections where electoral fraud  led to widespread demonstrations, the 2012 presidential elections were marred by a large number of irregularities although with a smaller impact on the electoral results. 
Compared to previous elections, it has been reported that in 2012 there was a major attempt to control the vote by such practices as massive voting using absentee voting certificates (AVCs)  or requiring employees to vote at their workplaces\footnote{http://www.gndem.org/node/3113}.
We observe greater statistical traces of pressure on voters in polling stations in the 2011 parliamentary elections where several types of fraud have been analyzed. 
A field experiment study carried out in Moscow estimates the size of fraud in voting shares for the ruling party  United Russia to be 11\% \cite{Enikolopova2012}.
A more recent research  has shown that the mix of electoral manipulation (electoral fraud, ballot stuffing, and voter pressure) used by incumbents varied across regions according to the competitive conditions. 
In particular, voter pressure was more common in competitive areas \cite{Harvey}.
To these findings we add the atypical skewing electoral behavior in small polling stations in order to understand the victory of the incumbent party. 

To conclude, the 2011 elections in Uganda require a totally different assessment. 
This country has enjoyed a more competitive landscape since the reintroduction of multiparty elections in 2005, but it has maintained the characteristics of electoral authoritarianism, key to keeping the National Resistance Movement (NRM) in power. 
In 2011, three practices common in past elections, also prevalent in Sub-Saharan Africa, are observed: intimidation, vote-buying and vote rigging \cite{Collier}.
In the context of the so-called monetization of elections, pressure on some voters was exerted through bribery (warning them of the consequences if they did not vote for the candidate who had bought their vote) and through the government development programs (threatening voters with the loss of benefits if they did not vote for the NRM)\footnote{http://thecommonwealth.org/sites/default/files/news-items/documents/Uganda-COG-Final-Report.pdf}.
In this case the statistical traces of bribery resemble those of voter rigging in the other two countries. The smaller the center, the larger the share of vote buying (Fig. 5). However, our method does not contemplate the difference between the mere purchasing of voting and the combination of buying and coercion.
In light of all these results, it is unclear to us which of the following two facts is more disturbing, namely (i) that  such large-scale distortions  of vote preferences keep recurring to the point that they may be outcome-determinative or (ii) that  such practices are so blatantly committed that they are hidden in plain statistical sight.

\section{Materials and Methods}

\subsection{Visualization of the SEFs}
We apply a convolution filter to the raw data of the election fingerprints, bottom row in Fig. 2, to obtain the smoothed contour visualizations for the SEFs, see Fig. 3.
Thereby we follow standard procedures by convoluting the raw data twice with a convolution kernel given by a ten-by-ten matrix with all entries being 0.01.

\subsection{Modified Thompson Tau Test}
This is a statistical test to identify outliers in a set of observations. 
The test has the advantage that it takes the observations' average and standard deviation into account.
Let $x$ be a vector of $n$ observations with average $\mbox{mean}(x)$ and standard deviation $\mbox{std}(x)$.
Furthermore, denote by $t_{\alpha/2}$ the $1-\tfrac{\alpha}{2}\times 100$ percentile of the Student's $t$ distribution with $n-2$ degrees of freedom.
One then computes the rejection threshold value $r = t_{\alpha/2} (n-1)/\sqrt{n(n-2+t^2_{\alpha/2})}$ and the vector $\Delta = |x-\mbox{mean}(x)|/\mbox{std}(x)$.
The test identifies the observation with the largest value $\Delta_i$ as outlier if $\Delta_i>r$.
If such an outlier exists, it is removed from $x$ and the test procedure is applied again on the remaining observations.
The test stops once all values in $\Delta$ are smaller than $r$.

\subsection{Outliers Removal}
We compute the 95\% confidence ellipse for the bivariate Gaussian distribution determined by the sample covariance matrix. Then we remove the $Z$-scores that lie outside of the ellipse. Although we are not assuming Gaussianity, in practice, this procedure corresponds to remove around 5\% of atypical electoral units of our case studies.

{\bf Funding.} R. Jimenez is supported by Spanish MINECO grant ECO2015-66593-P. M. Hidalgo  is supported by Spanish MINECO grant CSO2012-35852. P. Klimek is supported by the European Commission, FP7 project MULTIPLEX No. 317532.



\begin{thebibliography}{10}

\bibitem{Norris2014}
Norris P, Frank R and Martinez F (2014) Measuring Electoral Integrity around the world: A new dataset. {\em PS: Political Science \& Politics} 47:789--798.

\bibitem{Norris2014b}
Norris P (2014) {\em Why Electoral Integrity matters}. Cambridge University Press(New York).

\bibitem{JH}
Jim\'enez R and Hidalgo M (2014)
Forensic analysis of Venezuelan elections during the Ch\'avez presidency.
{\em PLoS ONE} 9(6):e100884.

\bibitem{Levin2013}
Levin I and Alvarez RM (2013)
Introduction to the Virtual Issue: Election Fraud and Electoral Integrity.
{\em Political Analysis} Virtual Issue. 

\bibitem{Mebane2008}
Mebane W (2008)
Election forensics: The second-digit Benford's law test and recent American presidential elections,
in {\em Election Fraud: Detecting and Deterring Electoral Manipulation}, eds Alvarez RM, Hall TE and Hyde SD (Brooking Press, Washington DC), pp 162--181. 

\bibitem{Pericchi2011}
Pericchi L and Torres D (2011)
Quick anomaly detection by the Newcomb-Benford Law, with applications to electoral processes data from the USA, Puerto Rico, and Venezuela.
{\em  Statist Sci} 26:513--527. 

 \bibitem{Cantu2011}
Cant\'u F and Saiegh SM (2011) 
Fraudulent Democracy? An Analysis of Argentina's Infamous Decade Using Supervised Machine Learning. 
{\em  Political Analysis} 19: 409--433.

\bibitem{Sacco2012}
Berber, B and Sacco, A (2012)
What the Numbers Say: A Digit-Based Test for Election Fraud. 
{\em  Political Analysis} 20:211--234.

\bibitem{Myakgov2009}
Myakgov M, Ordeshook PC, and Shaikin D (2009)
{\em The Forensics of Election Fraud}, Cambridge University Press. 

\bibitem{Klimek2012}
Klimek P, Yegorov Y, Hanel R, and Thurner S (2012)
Statistical detection of systematic election irregularities.
{\em Proc Natl Acad Sci USA} 109:16469--16473.

\bibitem{Levin2009}
Levin I, Cohn GA, Ordeshook PC, and Alvarez RM (2009) 
Detecting voter fraud in an electronic voting context: An analysis of the unlimited reelection vote in Venezuela,
in EVT/WOTE'09 Proceedings of the 2009 conference on Electronic voting technology/workshop on trustworthy elections, USENIX Association, Berkeley.

\bibitem{rj2011}
Jim\'enez R (2011) 
Forensic analysis of the Venezuelan recall referendum. 
{\em Statist Sci} 26:564--583.

\bibitem{Prado2011}
Prado R and Sans\'o B (2011)
The 2004 Venezuelan presidential recall referendum: Discrepancies between two exit polls and official results.
{\em Statist Sci} 26:502--512.

\bibitem{Hausmann2011}
Hausmann R. and Rigob\'on R. (2011) 
In search of the black swan: Analysis of the statistical evidence of fraud in Venezuela. 
{\em Statist Sci} 26:543--563.

\bibitem{Enikolopova2012}
Enikolopova R, Korovkina V, Petrovaa M, Sonina K, and Zakharovb A (2012)
Field experiment estimate of electoral fraud in Russian parliamentary elections.
{\em Proc Natl Acad Sci USA} 110:448--452.

\bibitem{Chatterjee2013}
Chatterjee A, Mitrovi\'c M and Fortunato S (2013) 
Universality in voting behavior: an empirical analysis. 
{\em Scientific Reports} 3:1049.

\bibitem{Borghesi}
Borghesi C, Raynal JC, Bouchaud JP (2012) Election turnout statistics in many countries: similarities, differences, and a diffusive field model for decision-making.
{\em PLoS ONE} 7(5):e36289. 

\bibitem{Mebane2011}
Mebane W (2011)
Comment on ÒBenfordÕs Law and the Detection of Election Fraud.
{\em  Political Analysis} 19: 269-272.

\bibitem{Freden}
Fred\'en A (2014)
Threshold Insurance Voting in PR Systems: A Study of VotersÕ Strategic Behavior in the 2010 Swedish General Election.
{\em Journal of Elections, Public Opinion and Parties} 24:473--492.

\bibitem{Franz}
Baringhaus L, Franz C (2004)
On a new multivariate two-sample test.
{\em Journal of Multivariate Analysis} 88:190--206.

\bibitem{Levitsky}
Levitsky S and Way LA (2010)
{\em Competitive Authoritarianism: Hybrid Regimes after the Cold War}. 
Cambridge University Press (New York). 

\bibitem{Schedler}
Schedler A (2002)
The Menu of Manipulation. 
{ \em Journal of Democracy} 13(2): 36-50. 

\bibitem{MH2009}
Hidalgo M (2009)
Hugo Ch\'avez's ``Petro-socialism''.
{\em Journal of Democracy} 20(2): 78-92. 

\bibitem{Harvey}
Harvey CJ (2016)
Changes in the menu of manipulation: Electoral fraud, ballot stuffing, and voter pressure in the 2011 Russian election.
{\em Electoral Studies} 41: 105-117.

\bibitem{Collier}
Collier P and Vicente PC (2012) 
Violence, bribery, and fraud: the political economy of elections in Sub-Saharan Africa. 
{\em Public Choice} 153(1): 117-147.

\bibitem{Girke}
Girke P and Kamp M (2011)
MuseveniÕs Uganda: Eternal Subscription for Power? 
{\em International Reports of the  Konrad-Adenauer-Stiftung} 5/2011.

\end{thebibliography}
\end{document}